\documentclass[aps,prb,twocolumn,floatfix,showpacs,superscriptaddress,floatfix]{revtex4}
\usepackage[english]{babel}
\usepackage{graphicx}
\usepackage{color}
\usepackage[utf8]{inputenc}
\usepackage{pstricks,pst-grad,color}
\usepackage{graphicx,amssymb}
\usepackage{amsmath}
\usepackage{amssymb}
\usepackage{bbold}
\usepackage{placeins}

\begin{document}

\title{Competition of striped magnetic order and partial Kondo screened state in the Kondo lattice model}

\begin{abstract}
We analyze the magnetically ordered phases in the Kondo lattice model on a square lattice around quarter filling using the dynamical mean field theory. 
We find that close to quarter filling besides the paramagnetic phase, at least three magnetic phases compete.
These phases are a ferromagnetic state, a partial Kondo screened state, which is a combination of charge order and magnetic order, and a striped magnetic phase, which is ferromagnetically ordered in one direction and antiferromagnetically in the other direction. We analyze  the spectral properties of these states and show that the PKS state is insulating with a flat band above the Fermi energy, while the other states are metallic.
Furthermore, we demonstrate that the energy gain by the Kondo effect is larger in the PKS state, while the striped magnetic state can gain more energy from the RKKY interaction. Thus, while the striped magnetic state is stable at weak coupling, the PKS state becomes stable at intermediate coupling.  
\end{abstract}

\author{Robert Peters}
\email[]{peters@scphys.kyoto-u.ac.jp}
\affiliation{Department of Physics, Kyoto University, Kyoto 606-8502, Japan}

\author{Norio Kawakami}
\affiliation{Department of Physics, Kyoto University, Kyoto 606-8502, Japan}

\newcommand{\1}{\mbox{1}\hspace{-0.25em}\mbox{l}}
\date{\today}



\maketitle
\section{introduction}
Heavy fermion materials are still at the center of research in strongly correlated electron systems. Due to the inclusion of partially filled $f$-electron orbitals  local magnetic moments are formed, which interact with the conduction electrons of the material.\cite{hewson1997} This coupling between localized magnetic moments and conduction electrons results in two different mechanisms which strongly affect the properties of the heavy fermion material. One mechanism is the Kondo effect, which leads to a singlet formation between the localized moments and the conduction electrons. The other  is the RKKY interaction, resulting in a long-range magnetic order of the localized moments.

Both mechanisms usually compete with each other, as best seen in the antiferromagnetic phase, where the magnetic phase vanishes via a second order phase transition when increasing the coupling strength between localized moments and the conduction electrons.\cite{doniach77,lacroix1979,fazekas1991,Assaad1999,Peters2007,Watanabe2007,Yamamoto2007,Martin2008,Watanabe2008,Lanata2008,Vojta2008,Yamamoto2008,Yamamoto2010b,Hoshino2010,Bodensiek2011,PhysRevB.82.245105,Asadzadeh2013,Si2014,Osolin2015,PhysRevB.92.075103} The common interpretation according to the Doniach phase diagram\cite{doniach77} is that the Kondo effect, whose energy scale depends exponentially on the coupling, gains more energy at strong coupling than the RKKY interaction. However, at intermediate coupling strengths in the antiferromagnetic phase, the Kondo effect and the magnetic order created by the RKKY interaction cooperate with each other to gain maximal energy in the ground state. This is demonstrated by the change of the Fermi surface within the antiferromagnetic phase from small surface at weak coupling to large Fermi surface at intermediate coupling.
 The cooperation of both mechanism results in a magnetic state due to the RKKY interaction with large Fermi surface due to the Kondo effect.\cite{Watanabe2007,Martin2008,Lanata2008,Watanabe2008,PhysRevB.82.245105,PhysRevB.92.075103} 

Depending on the filling of conduction electrons or the band structure, the interplay between these two mechanism changes. In the ferromagnetic state at low filling of the conduction electron band, this interplay results in a half-metallic state for a wide range of coupling strengths.\cite{Beach2008,yamamoto2010,PhysRevB.77.094419,PhysRevB.81.094420,Peters2012,Bercx2012,PhysRevB.86.165107,Golez2013,PhysRevB.87.134409,PhysRevB.92.094401,Kubo2015} The minority-spin conduction electrons are combined with the localized spins to form a Kondo insulator. Because this insulating state is only realized for one spin direction, this state has also been termed "spin-selective Kondo insulator".\cite{Peters2012}

Except for the antiferromagnetic state close to half-filling and the ferromagnetic state at low filling, the situation is less well understood.
Recently, it has been shown that exactly at quarter filling, $n=0.5$, the  interplay between Kondo effect and RKKY interaction, results in yet another surprising state: a charged ordered state.\cite{Peters2013,PhysRevLett.110.246401}
These calculations have shown that in this state  half of the lattice sites are magnetically ordered, while the rest of the lattice sites are nonmagnetic and have a different electron number. Thus, this state is a combination of a charge ordered state and magnetism. The charge order is thereby realized without the need of a nonlocal density-density interaction, as it is often necessary.
A recent Hartree-Fock study found a variety of different magnetic phases in the doped Kondo lattice model, but did not observe the PKS state.\cite{COSTA201774}

Motivated by these previous results, 
we want to ask the question whether other magnetic phases could be realized in the Kondo lattice model at intermediate fillings.
Indeed, we find that besides the charge ordered state another magnetic state, namely a striped magnetic state, exists around quarter filling. This striped magnetic state competes with the PKS state, which leads to a large region of phase coexistence. We analyze the properties of these phases and study the competition between them. 
By comparing the ground state energies of these states, we establish the detailed phase diagram of the Kondo lattice model on a square lattice around quarter filling including ferromagnetic, PKS, and striped magnetic order.

The remainder of this paper is organized as follows: after describing the model and method in the next section, we draw the magnetic phase diagram in section \ref{phase}, and show the spectral properties of these phases in section \ref{dynamics}. A conclusion finishes this paper.

\section{model and method}
\label{model}
In order to analyze the interplay between the Kondo effect and the RKKY interaction, we use the Kondo lattice model. Localized magnetic moments, originating from strongly correlated $f$-electrons, couple antiferromagnetically to conduction electrons.\cite{doniach77,lacroix1979,fazekas1991} The model reads
\begin{displaymath}
H=t\sum_{i,j,\sigma}c^\dagger_{i,\sigma}c_{j,\sigma}+J\sum_i\vec{S}\cdot\vec\sigma_{s_1s_2}c^\dagger_{i,s_1}c_{i,s_2},
\end{displaymath}
where the first term describes the hopping of conduction electrons on a square lattice with hopping amplitude $t$. The second term corresponds to a local antiferromagnetic coupling between the conduction electrons and the localized spins with coupling strength $J$.  This model is widely used for $f$-electron systems because it includes the Kondo effect and the RKKY interaction.

We use the dynamical mean field theory (DMFT)\cite{Metzner1989,Pruschke1995,Georges1996} to study this model. DMFT  maps the lattice model onto a quantum impurity model. Thus, this approximation includes local correlations and fluctuations but neglects non-local fluctuations. It has been widely used to describe $f$-electron systems and can well describe magnetism and heavy fermion behavior. In order to describe magnetic phases including the partial Kondo screened state, 
we use a $2\times 2$ sublattice and map each of the $4$ sites onto a separate impurity model. We include in our analysis a paramagnetic state (disordered), a ferromagnetic state, a Neel state, an A-type antiferromagnetic state and a PKS state. The Neel state did not yield converged solutions in the shown parameter regions. To figure out which magnetic state is the ground state at T=0, we compare the energies between all converged solutions. 

The resulting quantum impurity models are solved using the numerical renormalization group (NRG),\cite{wilson1975,Bulla2008} which is able to calculate real-frequency spectral functions and self-energies with high accuracy around the Fermi energy. The combination of DMFT and NRG has been frequently used to analyze $f$-electron systems and has been demonstrated to yield reliable results for the interplay between Kondo physics and magnetic order. \cite{Peters2006,Weichselbaum2007}

We note that by increasing the cluster size of the real-space DMFT calculations more complicated magnetic structures, such as long-range spin density waves, \cite{PhysRevB.89.155134,PhysRevB.92.075103} would possibly appear. However, we do not think that the results for the competition or cooperation between the Kondo effect and the RKKY interaction, as depicted by the magnetic phase diagram, will be strongly influenced by performing calculations for larger cluster sizes. A complete analysis of magnetic structures on larger cluster sizes is left for a future study.

\section{phase diagram}
\label{phase}
\begin{figure}[t]
\begin{center}
  \includegraphics[width=\linewidth]{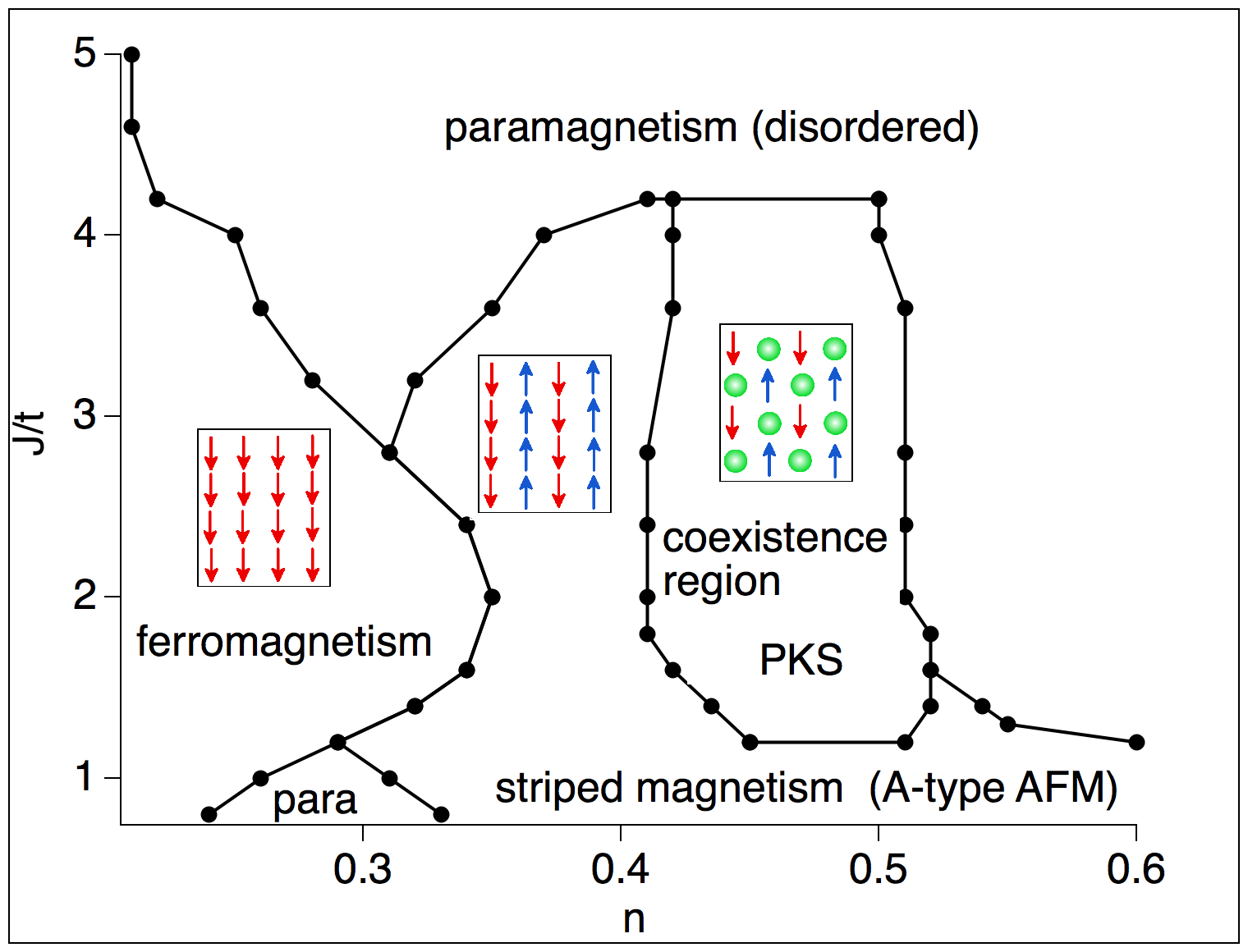}
\end{center}
\caption{Phase diagram around quarter filling including the paramagnetic phase, the ferromagnetic phase, the striped magnetic phase, and the partial Kondo screened  (PKS) state.  
\label{Fig1}}
\end{figure}

The main results of this paper are summarized in the phase diagram shown in Fig. \ref{Fig1}. We have identified four different phases in the vicinity of quarter filling for the Kondo lattice model on a square lattice. Besides the paramagnetic (nonmagnetic) and the ferromagnetic phase, we find the partial Kondo screened (PKS) state and a striped magnetic state. The striped magnetic state,  which corresponds to an A-type antiferronagnetic state on the square lattice,  is antiferromagnetically ordered in one direction and ferromagnetically ordered in the other direction.
The PKS phase, which was reported by \textcite{PhysRevLett.110.246401}, is a combination of charge order and magnetic order. On a $2\times 2$ cluster, two diagonal sites form a magnetic state, while the other two sites are nonmagnetic. The magnetic sites and nonmagnetic sites have different electron fillings. Combining  all sites, the electron filling is in average  $n=0.5$ per site.
The spin configuration of each magnetic state is shown in Fig. \ref{Fig1}. 
\begin{figure*}[t]
\begin{center}
    \includegraphics[width=0.32\linewidth]{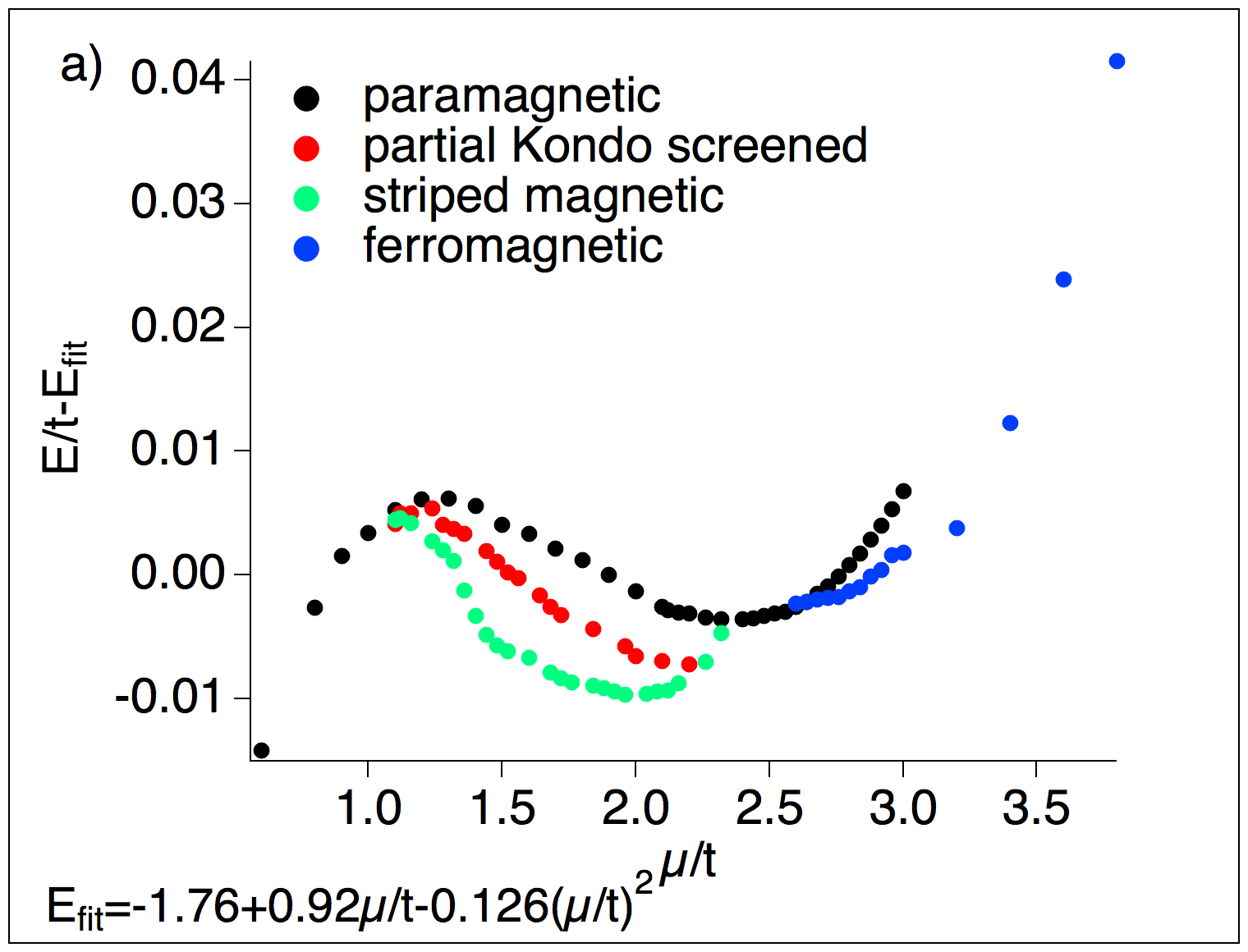}
  \includegraphics[width=0.32\linewidth]{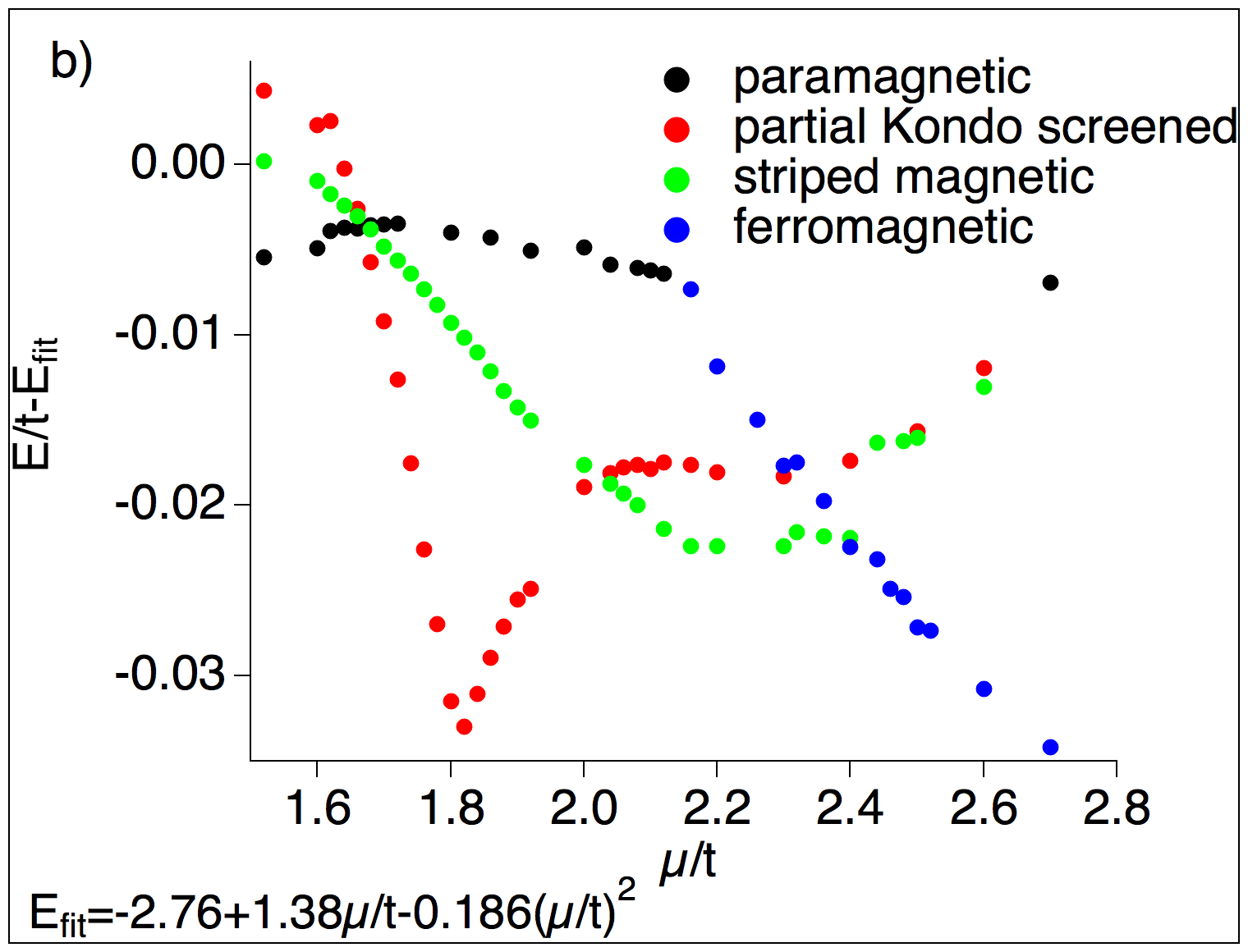}
    \includegraphics[width=0.32\linewidth]{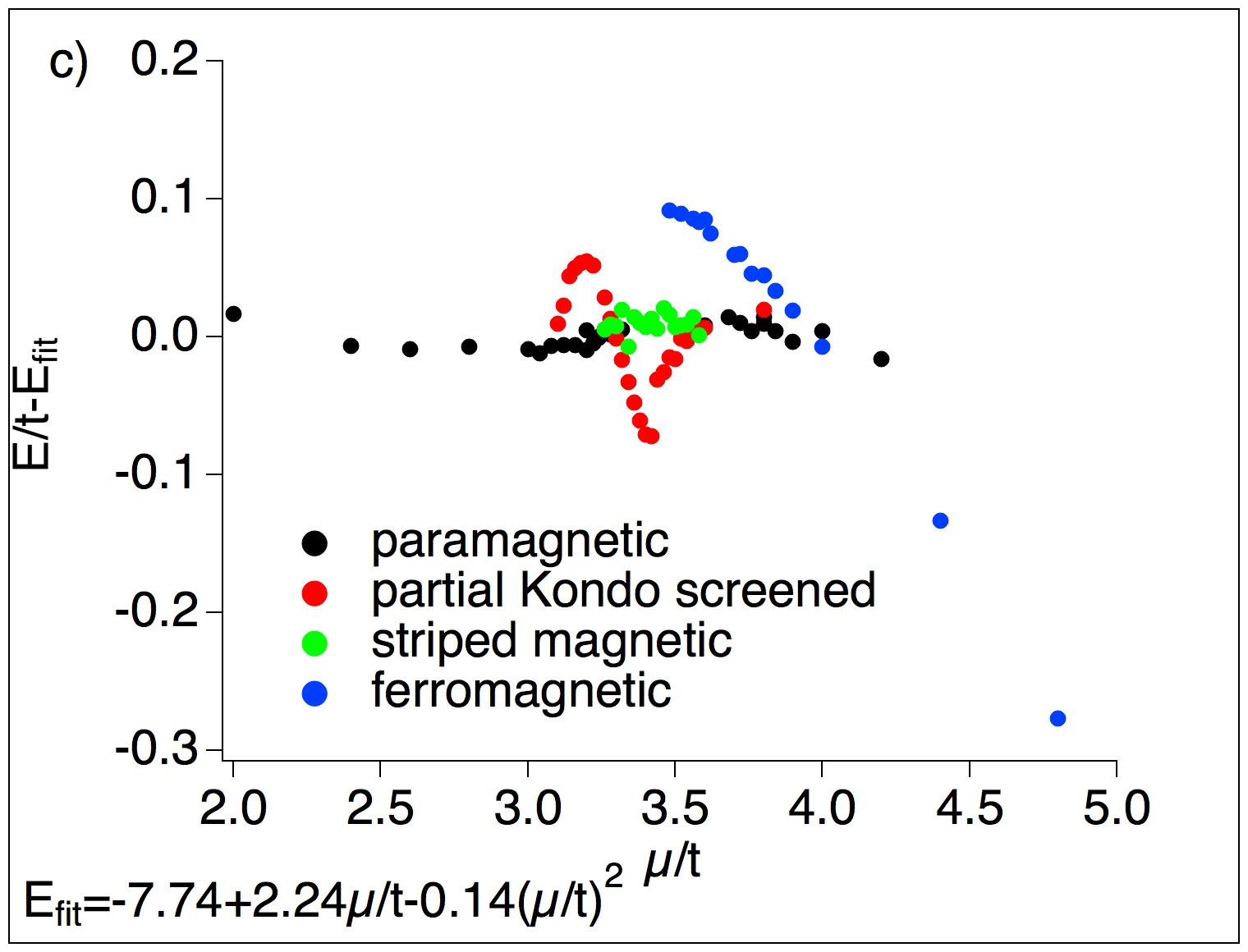}
\end{center}
\caption{Energy comparison between striped magnetic state, PKS state, and ferromagnetic state for different chemical potentials. The panels show the results for $J/t=0.8$, $J/t=1.6$, and $J/t=4$ from left to right. We have subtracted a polynomial fit to the paramagnetic state from each energy for a better visualization of the comparison. 
\label{Fig2}}
\end{figure*}

We start our explanations of the phase diagram at weak coupling. 
For coupling strengths $J/t<1.2$, the striped magnetic state is the energetically stable state for $0.32<n<0.65$ and changes into a paramagnetic state approximately at $n=0.32$. As we will show below, the PKS state has for all chemical potentials the higher energy for weak coupling strengths.
For filling $n<0.26$, the Kondo lattice forms a ferromagnetic state for $J/t=1$. These results are consistent with previous findings,\cite{PhysRevB.92.075103}
where it was shown that the small-Fermi-surface SDW phase changes via a first order transition below $n<0.7$ into another magnetic state, which we have now identified as the striped magnetic state shown in Fig. \ref{Fig1}. 

For coupling strengths $J/t>1.2$, the situation becomes more complicated. At filling $n\approx 0.6$,  the paramagnetic state becomes energetically stable. 
However, around quarter filling, $n=0.5$, several phases coexist. Besides the paramagnetic state, which is the energetically stable state for $n>0.51$, the striped magnetic state and the PKS state coexist. 
Although the PKS state is the energetically lowest state for a wide range of chemical potential, it can only exists exactly at quarter filling. The electron filling of the paramagnetic state and the striped magnetic state, on the other hand, change in this coexistence region.  We will explain this region of the phase diagram in more detail below.

For electron filling  $n\approx 0.4$, the striped magnetic phase becomes the energetically lowest state and the PKS state cannot be stabilized anymore. Below $J/t=2.8$, the striped magnetic state changes via a first order transition to the ferromagnetic state at approximately $0.3<n<0.35$. Above $J/t=2.8$ the striped magnetic state vanishes when decreasing the electron filling via a second order phase transition  before the ferromagnetic state becomes stable.  Around quarter filling, the PKS state is energetically stable for coupling strengths up to $J/t=4.2$. For $J/t>4.2$, besides the ferromagnetic state at low filling, we only find the paramagnetic state, which covers most of the phase diagram. 
 
In order to understand this phase diagram, we compare  the energies of the states for different chemical potentials and interaction strengths in Fig. \ref{Fig2}.
For a better visualization, we have subtracted a polynomial fit to the energy of the paramagnetic state, $E_{fit}$, from all our results. Without subtracting this fitting energy, the energies are monotonically increasing with increasing chemical potential $\mu/t$. However, due to the wide range of chemical potentials shown in these figures, differences between different states are not clearly visible without the subtraction. 

Figure \ref{Fig2}(a) shows the energies for coupling strength $J/t=0.8$. For chemical potentials $\mu/t>1.1$ the striped magnetic state is the energetically lowest and thus stable state. Increasing the chemical potential, this state vanishes via a second order phase transition, and the paramagnetic state becomes stable. Finally for $\mu/t>2.4$  the ferromagnetic state becomes the ground state. Although the PKS state can be stabilized for this interaction strength, it never becomes the energetically lowest state. 
 
Figure \ref{Fig2}(b) shows the results for $J/t=1.6$. We see that while for small chemical potentials the paramagnetic state is the energetically lowest state, for $1.7<\mu/t<2.0$ the PKS state becomes the ground state. Although the PKS state is the ground state for a wide range of chemical potentials, the electron filling of this state is fixed to $n=0.5$, thus $\frac{dn}{d\mu}=0$. 
 For the same chemical potentials also the paramagnetic state and the striped magnetic state can be stabilized. The electron filling of the paramagnetic state and the striped magnetic state decrease when increasing the chemical potential. The electron filling of these states vary approximately from $n=0.51$ to $n=0.41$ in the range of chemical potentials where the PKS state is the energetically lowest state. 
  This coexistence of different states with different electron filling will lead in an experimental situation to phase separation between these phases. If we dope holes into a PKS state with $n=0.5$, regions with striped magnetic state will appear, because while the striped magnetic state can be doped, the PKS state only exists at $n=0.5$.
On the other hand, starting with a striped magnetic state with filling $n=0.4$ and dope electrons into it, the material might form regions of PKS state.

Figure \ref{Fig2}(c) shows the energy comparison for $J/t=4$. For this interaction strength, the energy of the striped magnetic state and the paramagnetic state are very close. Around quarter filling we find a similar behavior as for intermediate coupling strengths and the PKS state is the energetically lowest state. This state still coexists with the striped magnetic state. However, with increasing the chemical potential the striped magnetic state vanishes via a second order phase transition and the ground state becomes paramagnetic at approximately $n=0.4$. The paramagnetic state is the ground state of the Kondo lattice model until reaching the ferromagnetic state at $n\approx 0.3$.

Taking into account this energy comparison for all calculated coupling strengths, we end up with the phase diagram in Fig. \ref{Fig1}. The coexistence region around $n=0.5$ implies that the PKS state has the lowest energy in this region, but can be stabilized only at quarter filling. Therefore, for electron filling different from $n=0.5$, phase separation between the PKS state and the striped magnetic state can occur.

\begin{figure}[t]
\begin{center}
  \includegraphics[width=\linewidth]{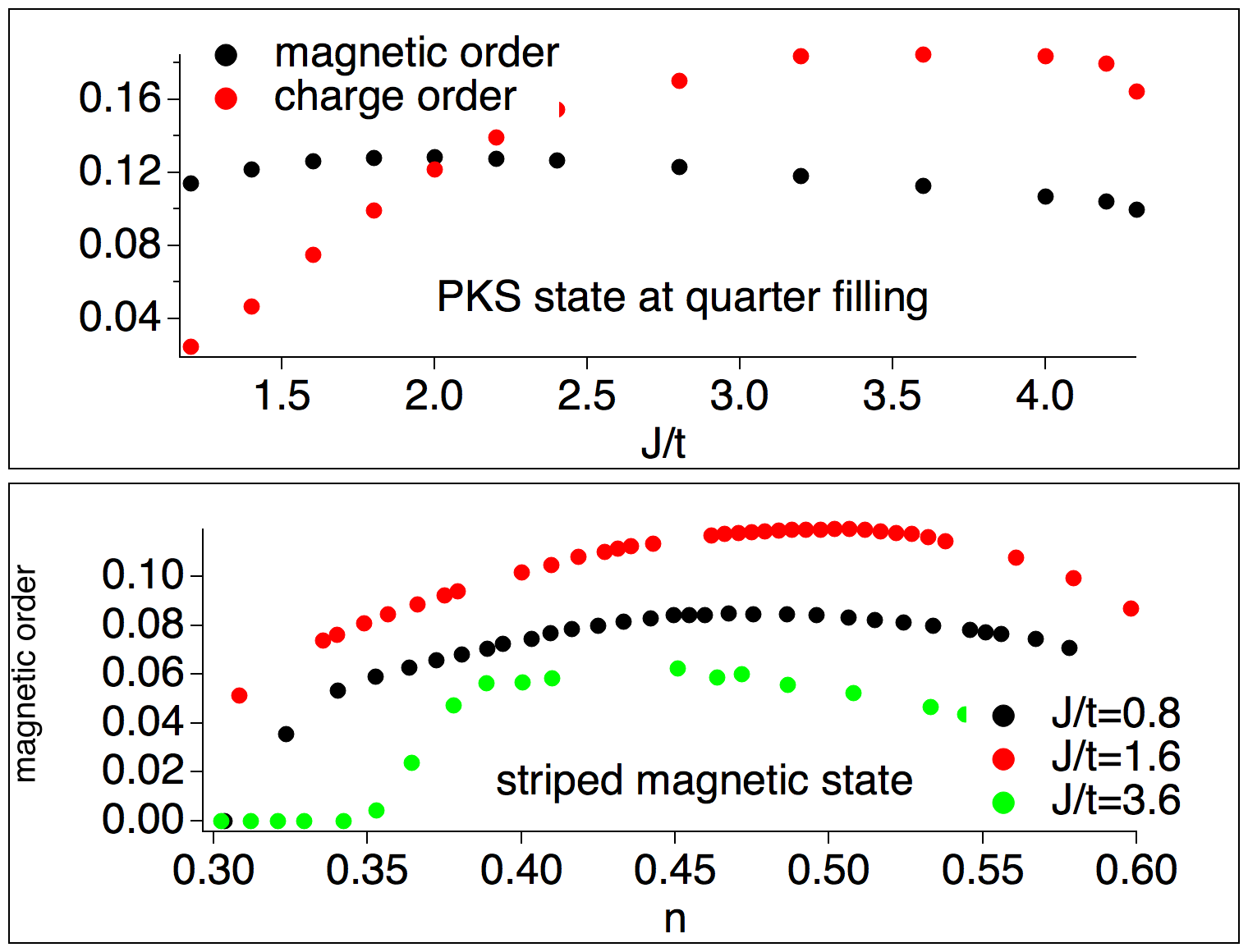}
\end{center}
\caption{Evolution of the order parameter for the PKS state and the striped magnetic state. Top: magnetic order and charge order for the PKS state for different interaction strengths at quarter filling. Bottom: Magnetic order in the striped magnetic phase for different interaction strengths and electron numbers.
\label{Fig3}}
\end{figure}

 Finally, in Fig. \ref{Fig3} we show the order parameter of the PKS state and the striped magnetic state. The top panel shows the magnetic order and the charge order of the PKS state at quarter filling for different coupling strengths. All phase transitions into the PKS state are of first order, showing a jump of the order parameter.
While for coupling strengths below $J/t=1.2$, the striped antiferromagnetic state becomes energetically stable at quarter filling, for coupling strengths larger than $J/t=4.2$ no magnetic state can be stabilized at quarter filling. This phase transition agrees with the observations made by variational Monte Carlo.\cite{PhysRevLett.110.246401}

\begin{figure}[t]
\begin{center}
  \includegraphics[width=\linewidth]{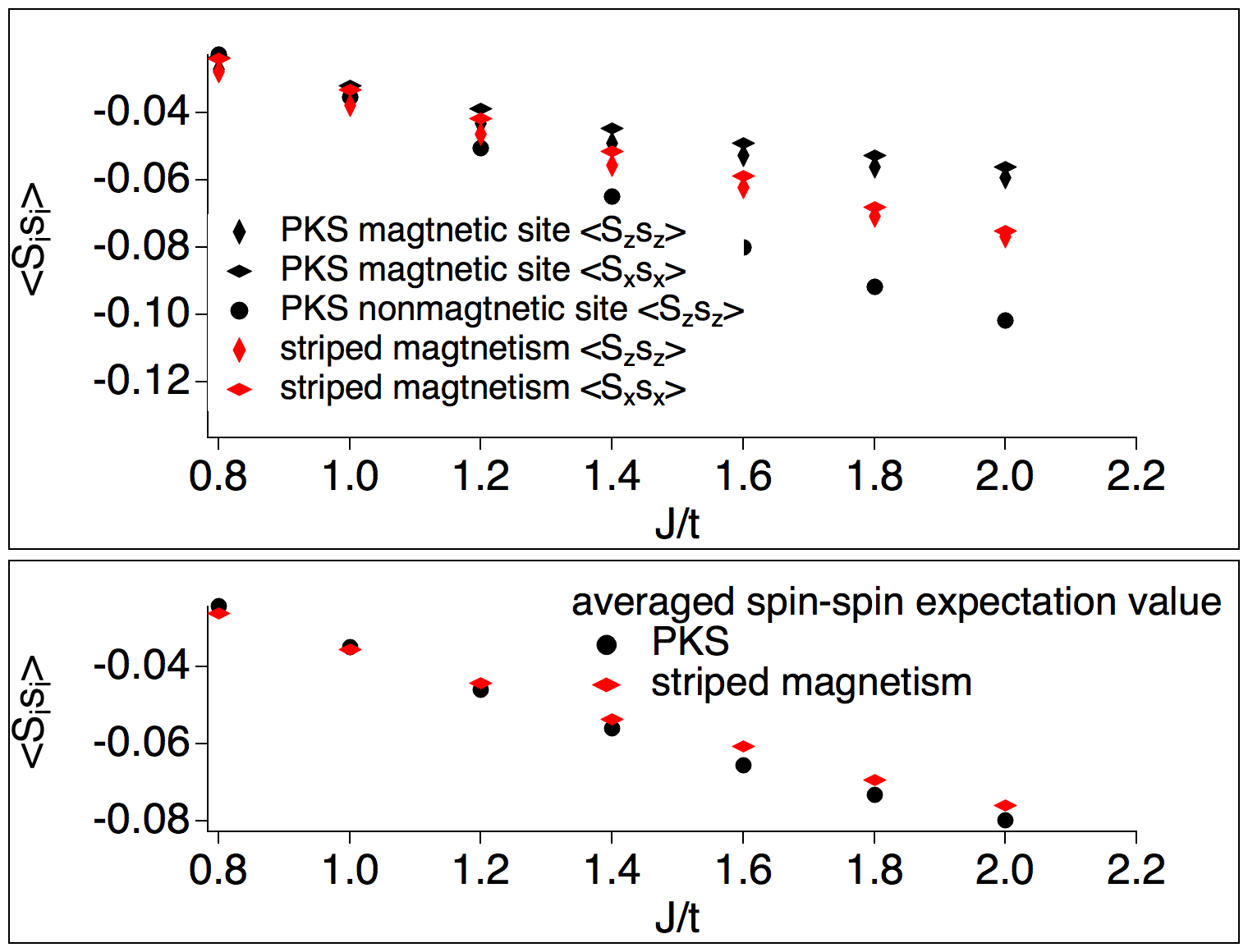}
  \end{center}
\caption{Local Spin-Spin correlation at quarter filling, $n=0.5$, for the PKS state and the striped magnetic state. The bottom panel shows the averaged expectation value, which is proportional to the energy by the spin-spin interaction.
\label{Fig4}}
\end{figure}

\begin{figure*}[t]
\begin{center}
  \includegraphics[width=0.3\linewidth]{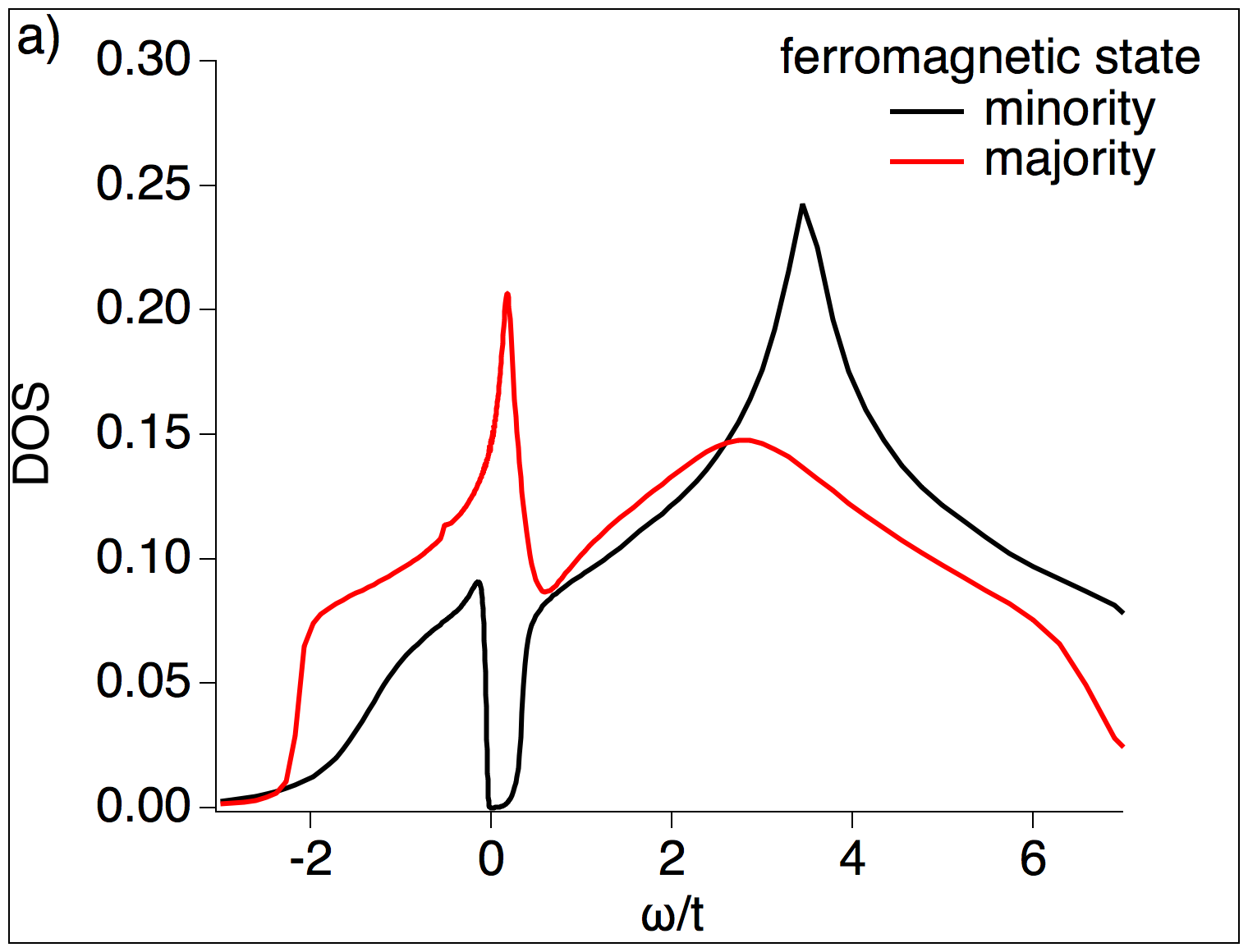}
    \includegraphics[width=0.3\linewidth]{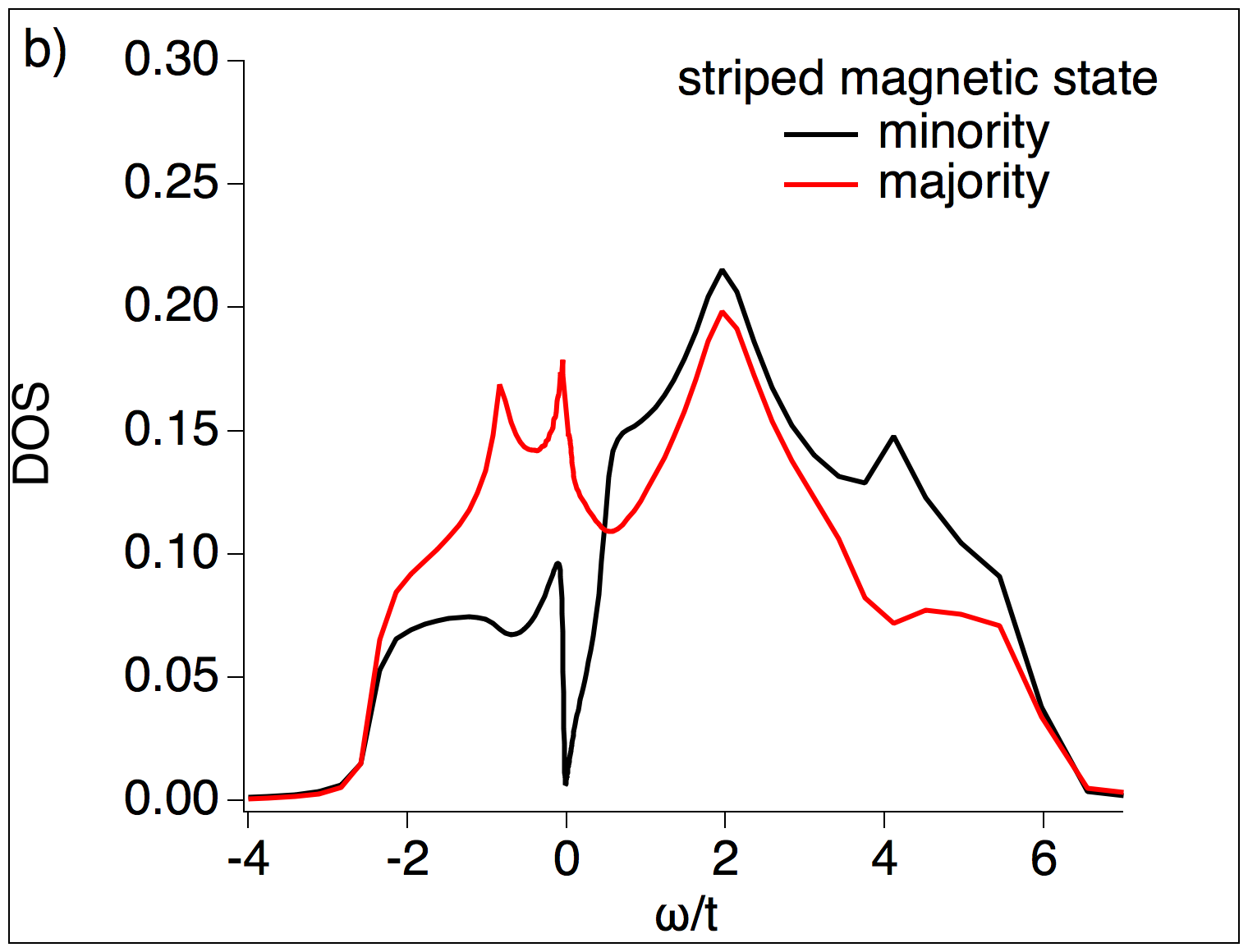}
        \includegraphics[width=0.3\linewidth]{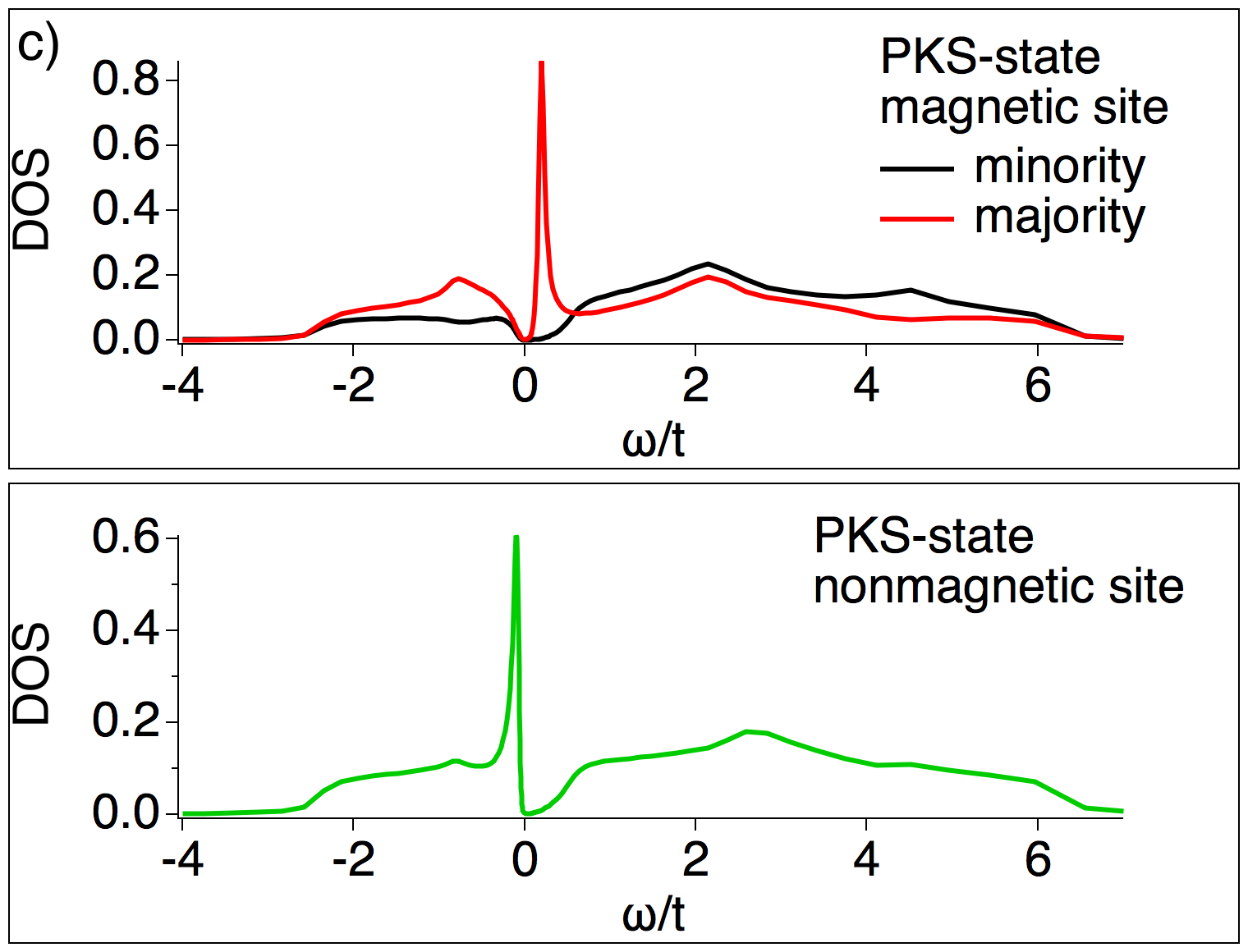}
                    \includegraphics[width=0.24\linewidth]{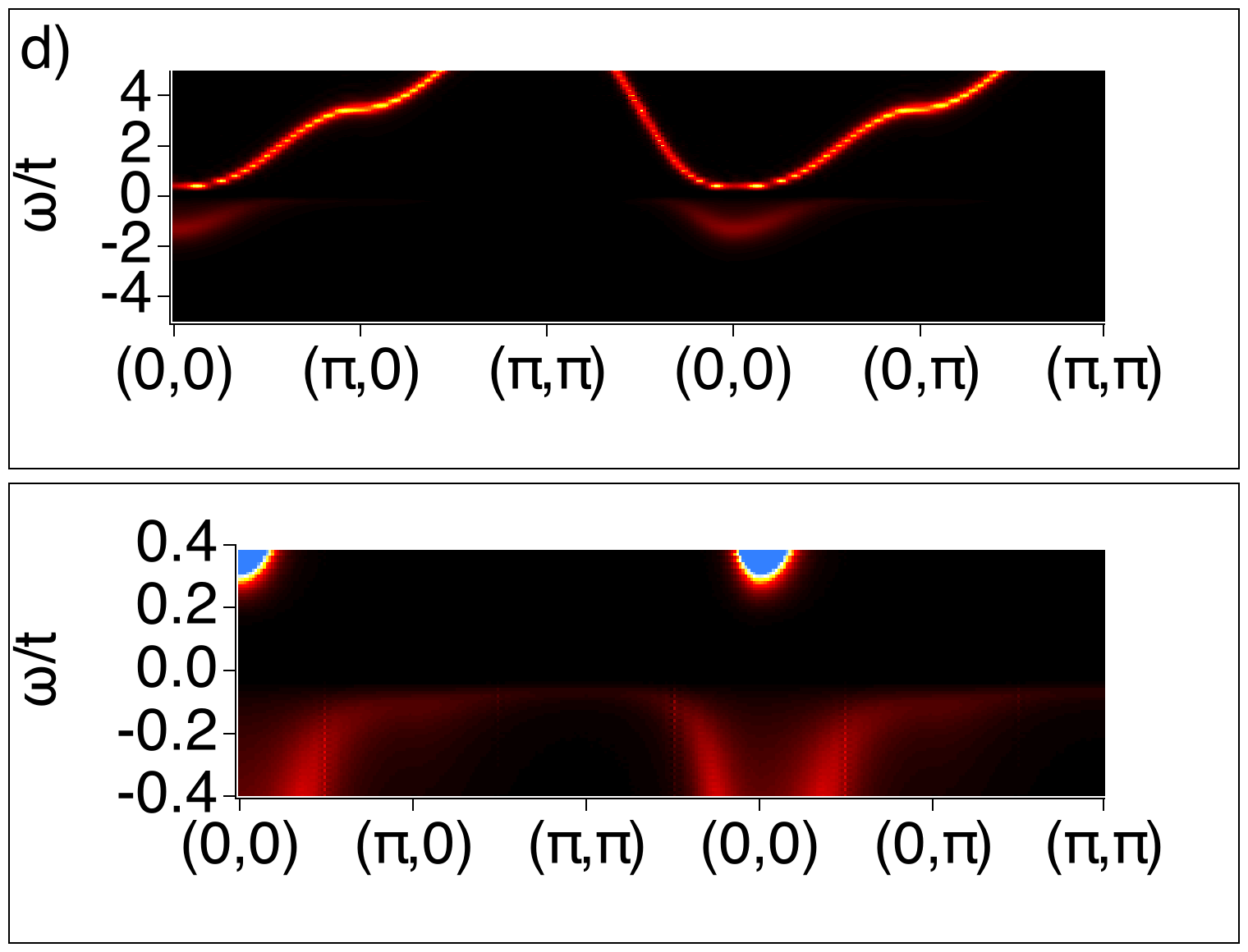}
                \includegraphics[width=0.24\linewidth]{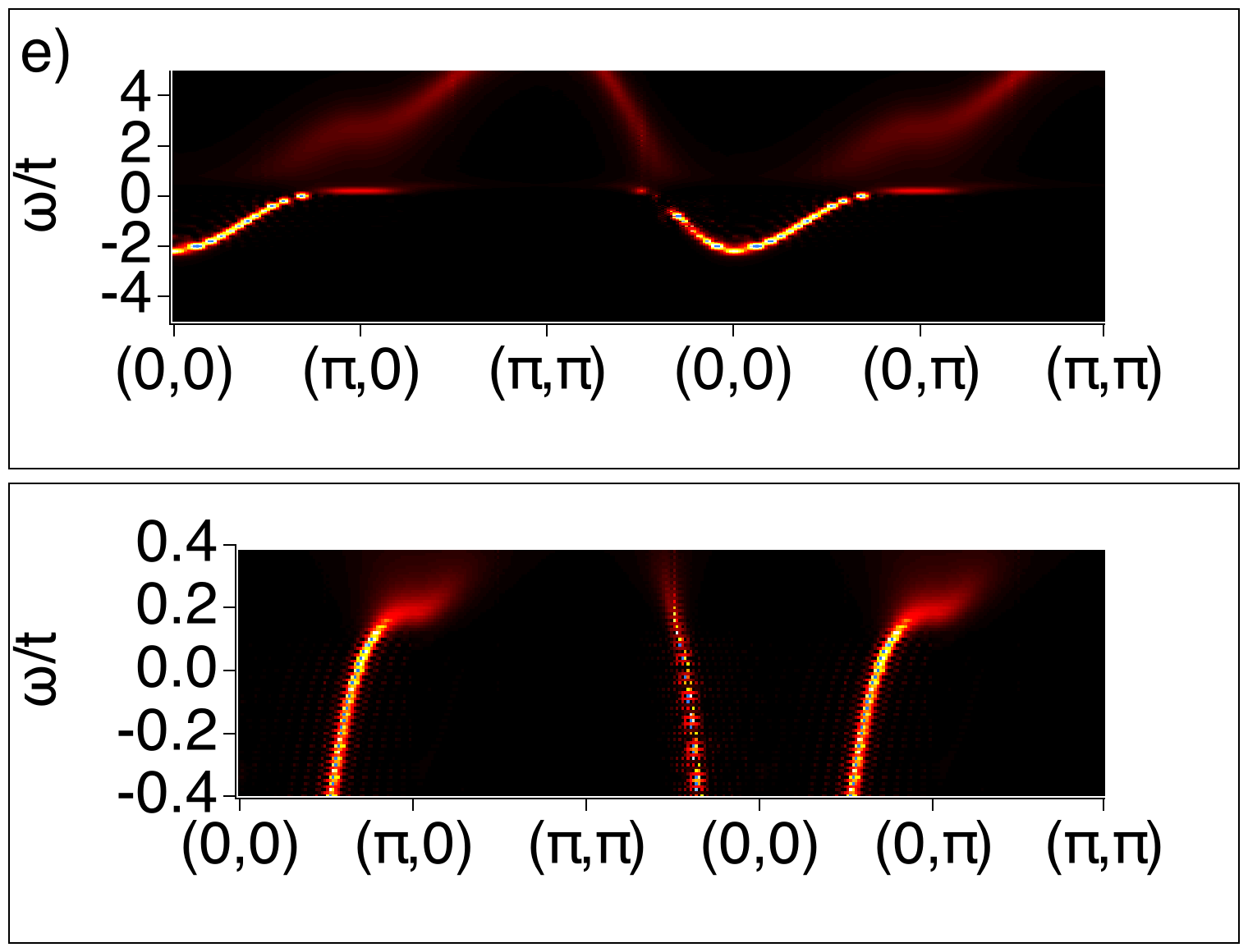}
            \includegraphics[width=0.24\linewidth]{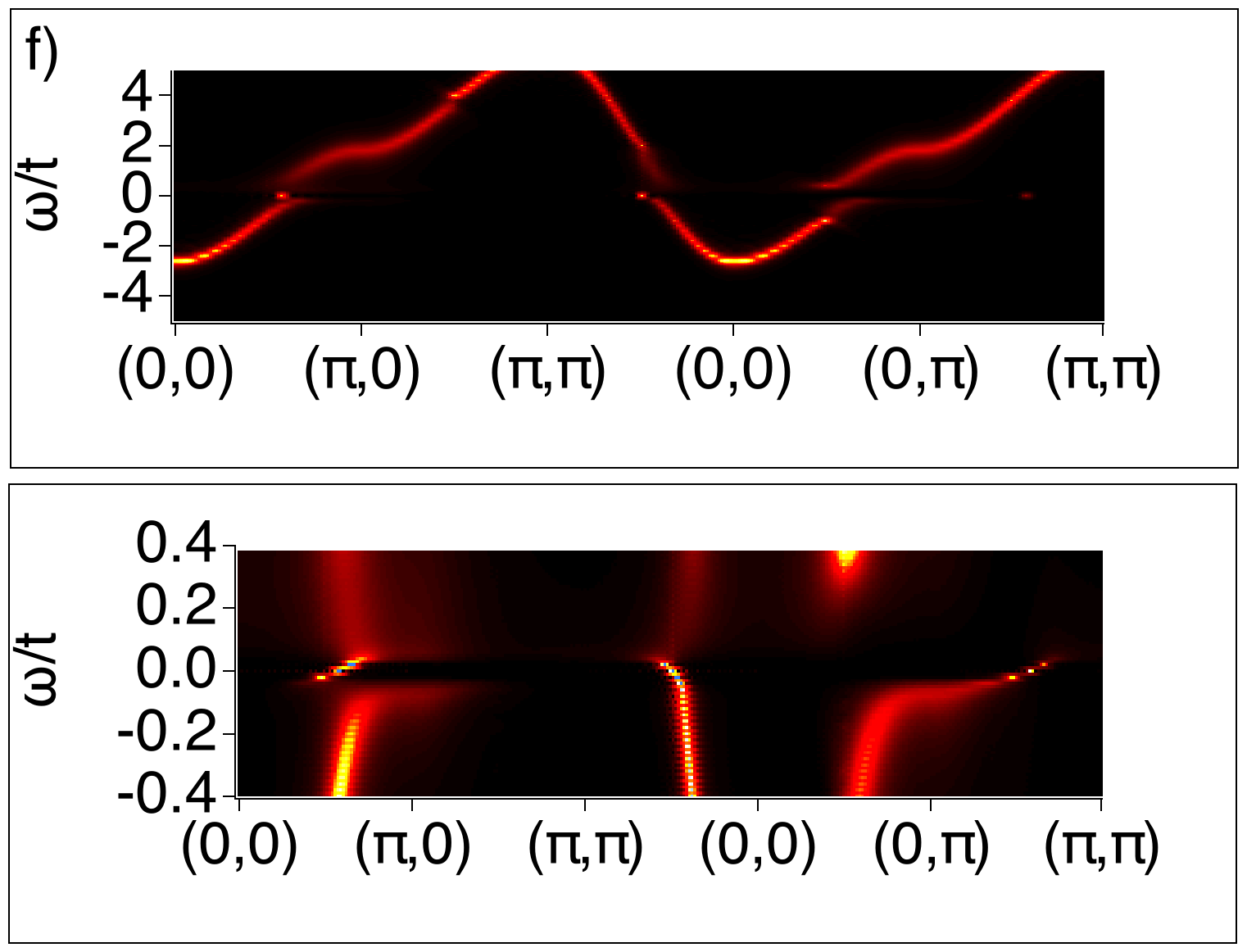}
                \includegraphics[width=0.24\linewidth]{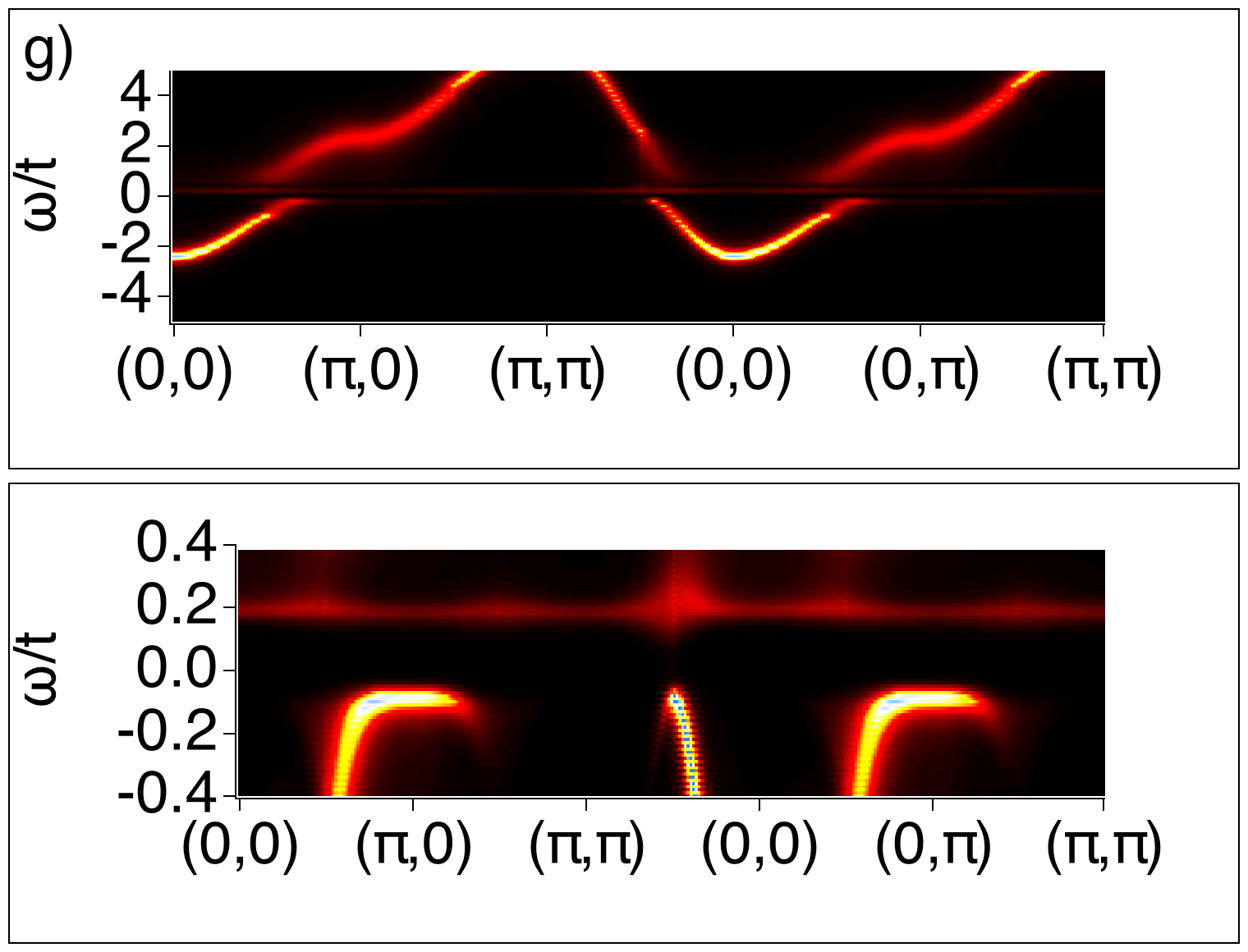}
\end{center}
\caption{Local density of states and momentum resolved spectrum for the different states. The upper panels show the local density of states of a) the ferromagnetic state, for $J/t=2.4$ and $n=0.3$, b) the striped magnetic state for $J/t=1.6$ and $n=0.48$, and c) the PKS state for $J/t=1.6$ and $n=0.5$. The lower panels show the momentum resolved spectra of d) the minority electrons in the ferromagnetic state, e) the majority electrons in the ferromagnetic state, f) the striped magnetic state, and g) the PKS state for the same parameters as the DOS. 
\label{Fig5}}
\end{figure*}
The bottom panel of Fig. \ref{Fig3} shows the order parameter of the striped magnetic state for different coupling strengths and electron filling. By starting the DMFT calculation with an converged solution of a striped magnetic state of a nearby interaction strength and chemical potential, one can force the Kondo lattice model into this state, although it might not be the energetically lowest state.
Thus, by forcing the system into this state, the striped magnetic state can be stabilized for electron filling  $0.4<n<0.5$ where the magnetic order vanishes continuously. Increasing the coupling strength above $J/t=4$ the striped magnetic state vanishes continuously for any particle number.  
However, the energy comparison  above has shown that the striped magnetic state is only the ground state in the whole region of chemical potentials for $J/t<1$.  Above
$J/t>1.2$ the striped magnetic state coexists with the PKS state, which has the lower energy around quarter filling. Thus, we find a first order transition and phase coexistence around quarter filling.
Furthermore, for intermediate coupling strengths, $1.2<J/t<3$, the ferromagnetic state becomes the ground state at low filling before the striped magnetic state vanishes, which results in a first order transition between the ferromagnetic state and the striped magnetic state.

As shown in Figs. \ref{Fig1} and \ref{Fig2}, all these states compete with each other. The ground state energy is the sum of the kinetic energy and the local interaction energy created by the antiferromagnetic spin-spin interaction, which leads to a singlet formation. 
The strength of the resulting energy gain by the singlet formation differs for each phase. Generally, the energy gain by the singlet formation will behave for weak coupling like the Kondo energy. Thus, it will depend exponentially on the coupling strength and the density of states of the noninteracting model at the Fermi energy, which results in a dependence of the energy gain on the electron filling. For large coupling, on the other hand, the energy gain by the singlet formation will saturate and  becomes linear in the coupling strength. We note that the main energy difference between different states at the same electron filling comes from the interaction energy, which is proportional to the local spin-spin correlation. The kinetic energy plays a minor role.

In the upper panel of Fig. \ref{Fig4}, we show the correlations $\langle S_z s_z\rangle$ and $\langle S_x s_x\rangle=\langle S_y s_y\rangle$  of the PKS state and the striped magnetic state at quarter filling. The magnetic order is realized in $z$-direction. Therefore, $\langle S_z s_z\rangle\neq \langle S_x s_x\rangle$ for the magnetically ordered sites. For the nonmagnetic sites of the PKS state, the spin correlations are identical in all directions. 

For the PKS state, 
magnetically ordered sites and nonmagnetic sites have different electron filling.
Due to this charge order, the PKS state gains more energy by singlet formation than a paramagnetic state with homogenous charge distribution.\cite{Peters2013} On the other hand, by comparing the spin-spin correlations  between the PKS state and the striped magnetic state, the striped magnetic state gains more energy on the  magnetic sites.  
The average spin-spin correlation, which is proportional to the average interaction energy, is shown in the bottom panel of Fig. \ref{Fig4}. We see that the energy gain of the striped magnetic phase is larger for weak coupling, where the energy gain by the singlet formation due to the Kondo effect is small. When the Kondo effect becomes large, the PKS state gains more energy by singlet formation. Finally, for large coupling strength, the spin-spin correlation saturates. Consequently, the paramagnetic state approaches the same energy, and finally there is no energy gain by forming an ordered state.

 Our finding that the PKS state is realized at intermediate coupling strengths through a gain of Kondo energy, while the striped magnetic state exists at quarter filling for weak coupling, agrees with the fact that the striped magnetic state can be observed in the ferromagnetically coupled Kondo lattice model.\cite{PhysRevB.79.064411} Because the Kondo effect is absent in the ferromagnetically coupled Kondo lattice model, the striped magnetic state (or A-type antiferromagnetic state) is stable in a large region of the phase diagram.

\section{dynamical properties}
\label{dynamics}
An advantage of using the combination of DMFT and NRG is the ability to calculate spectral functions on the real-frequency axis without the need of analytical continuation. 
We show local and momentum resolved spectra of all magnetic states in Fig. \ref{Fig5}. 

We start our analysis of the spectral functions with the ferromagnetic state, which we show here for completeness.
The ferromagnetic state in the Kondo lattice model 
forms a gap in the DOS of the minority electrons at the Fermi energy due to partial Kondo screening.\cite{Peters2012} The gap in the DOS of the minority electrons, which exists for a wide range of interaction values and is independent of the filling, is clearly visible in the local DOS in Fig. \ref{Fig5}(a) and the momentum resolved spectrum in Fig. \ref{Fig5}(d). The DOS of the majority electrons, shown in Fig. \ref{Fig5}(a) and (e), on the other hand, exhibits a peak close to the Fermi energy.

\begin{figure}[t]
\begin{center}
  \includegraphics[width=\linewidth]{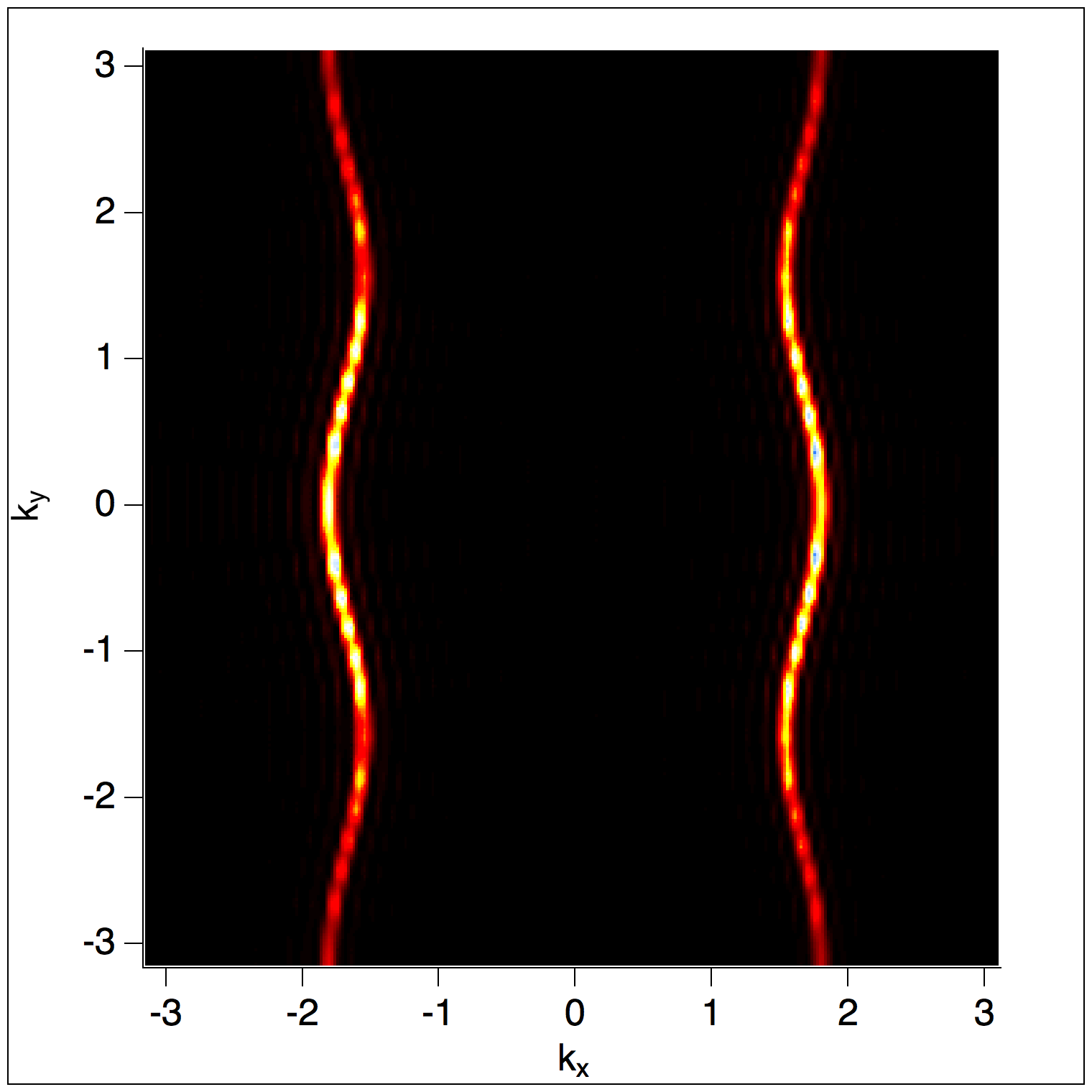}
  \end{center}
\caption{Fermi surface of the striped magnetic state for $J/t=1.6$ and $n=0.48$.
\label{Fig6}}
\end{figure}

Looking at the local DOS of the striped magnetic state in Fig. \ref{Fig5}(b), it is surprising to find a dip in the DOS of the minority electrons of each site, while the majority electrons have a peak at the Fermi energy. Thus, the effect of the partial Kondo screening, which leads to a gap in the minority conduction electrons for the ferromagnetic state, can be observed in the striped magnetic state, which exhibits ferromagnetic order in one direction. 
However, due to the antiferromagnetic order in the other direction, majority and minority conduction electrons exchange. 
One can thus expect that this dip at the Fermi energy of the minority electrons is not a full gap.
 Furthermore, looking at the expectation values of electron filling, magnetization, and polarization of the local moment, we cannot find any commensurability condition as in the ferromagnetic state. An analysis of the  momentum resolved spectrum, Fig. \ref{Fig5}(f), reveals that there are bands crossing the Fermi energy. 
 As can be expected from the symmetry of this state, in one direction antiferromagnetically ordered and in the other direction ferromagnetically ordered, the momentum resolved spectral function is asymmetric when interchanging $k_x$ and $k_y$. While a band crosses the Fermi energy between $(k_x,k_y)=(0,0)$ and $(\pi,0)$, there is no such band between $(k_x,k_y)=(0,0)$ and $(0,\pi)$. We find a one-dimensional like Fermi surface, shown in Fig. \ref{Fig6}, which will lead to very asymmetric transport properties at low temperatures.

Finally, we show the DOS of the PKS state in Fig. \ref{Fig5}(c) and (g).  The PKS state, which is realized only at quarter filling, is an insulating state exhibiting a gap at the Fermi energy. The PKS state, which is depicted in the phase diagram in Fig. \ref{Fig1}, combines $4$ sites, which results in a commensurate filling and explains the resulting gap. The momentum resolved spectrum in Fig. \ref{Fig5}(g) shows a nearly momentum independent flat band above the gap. An analysis of the local and spin-resolved DOS yields that this band is created by the magnetic sites. The nonmagnetic site, on the other hand, shows a peak below the Fermi energy. 
\section{conclusions}
\label{discussion}

We have analyzed magnetic phases in the Kondo lattice model on a square lattice around quarter filling. 
Besides the known phases, such as antiferromagnetic phase, ferromagnetic phase, and PKS state, we have demonstrated the existence of another magnetic state. i.e. the striped magnetic state. Based on these calculations, we have established a detailed phase diagram.
Furthermore, we have shown that the Kondo effect plays an important role in all these phases. The ferromagnetic phase, as a spin-selective Kondo insulator, stabilizes a magnetic polarization so that one spin-direction can gain the maximum possible Kondo energy, which leads to a gap at the Fermi energy of the minority electrons. The striped magnetic state shows similar features. However, because the RKKY interaction  favors a different ordering vector, a combination of ferromagnetic state in one direction and an antiferromagnetic state in the other direction is realized. Due to this order, a one-dimensional Fermi surface is formed. 
Finally, the PKS state combines charge order and magnetic order, which leads to an increase in the energy gain of the Kondo energy, but a loss of the RKKY energy. 
Due to a commensurate filling, this state is insulating.
The competition between the PKS state and the striped magnetic state results in a wide coexistence region around quarter filling. 
The shown phase diagram elucidates the interesting physics which emerges from the interplay between the RKKY interaction and Kondo effect depending on the electron filling.

\begin{acknowledgments}
This work was partly supported by a Grant-in-Aid for Scientific Research  
on Innovative Areas (JSPS KAKENHI Grant No. JP15H05855) and also JSPS  
KAKENHI (No. JP16K05501).
\end{acknowledgments}

\bibliography{paper}
\end{document}